\documentclass[fleqn,10pt]{wlscirep}
\usepackage[utf8]{inputenc}
\usepackage[T1]{fontenc}
\DeclareUnicodeCharacter{2212}{\textendash}
\usepackage[displaymath, mathlines]{lineno}
\linenumbers
\title{Universal sublinear resistivity in vanadium kagome materials hosting charge density waves}

\author[1,*]{Shirin Mozaffari}
\author[1]{William R. Meier}
\author[1]{Richa P. Madhogaria}
\author[2]{Nikolai Peshcherenko}
\author[3]{Seoung-Hun Kang}
\author[4]{John W. Villanova}
\author[5]{Hasitha W. Suriya Arachchige}
\author[6]{Guoxin Zheng}
\author[6]{Yuan Zhu}
\author[6]{Kuan-Wen Chen}
\author[6]{Kaila Jenkins}
\author[6]{Dechen Zhang}
\author[6]{Aaron Chan}
\author[6]{Lu Li}
\author[4]{Mina Yoon}
\author[5,7]{Yang Zhang}
\author[1,3,5,*]{David G. Mandrus}
\affil[1]{Department of Materials Sciences and Engineering, University of Tennessee, Knoxville, TN 37996, USA}
\affil[2]{Max Planck Institute for Chemical Physics of Solids, 01187, Dresden, Germany}
\affil[3]{Materials Science and Technology Division, Oak Ridge National Laboratory, Oak Ridge, Tennessee 37831, USA}
\affil[4]{Center for Nanophase Materials Sciences, Oak Ridge National Laboratory, Oak Ridge, Tennessee 37831, USA}
\affil[5]{Department of Physics and Astronomy, University of Tennessee, Knoxville, TN 37996, USA}
\affil[6]{Department of Physics, University of Michigan, Ann Arbor, MI 48109, USA}
\affil[7]{Min H. Kao Department of Electrical Engineering and Computer Science, University of Tennessee, Knoxville, Tennessee 37996, USA}

\affil[*]{corresponding: smozaff1@utk.edu, dmandrus@utk.edu}

\begin{abstract}
	The recent discovery of a charge density (CDW) state in ScV$_6$Sn$_6$ at $T_{\textrm{CDW}}$ = 91 K offers new opportunities to understand the origins of electronic instabilities in topological kagome systems. By comparing to the isostructural non-CDW compound LuV$_6$Sn$_6$, we unravel interesting electrical transport properties in ScV$_6$Sn$_6$, above and below the charge ordering temperature. We observed that by applying a magnetic field along the $a$ axis, the temperature behavior of the longitudinal resistivity in ScV$_6$Sn$_6$ changes from metal-like to insulator-like above the CDW transition. We show that in the charge ordered state ScV$_6$Sn$_6$ follows the Fermi liquid behavior while above that, it transforms into a non-Fermi liquid phase in which the resistivity varies sublinearly over a broad temperature range. The sublinear resistivity, which scales by $T^{3/5}$ is a common feature among other vanadium-containing kagome compounds exhibiting CDW states such as KV$_3$Sb$_5$, RbV$_3$Sb$_5$, and CsV$_3$Sb$_5$. By contrast, the non-Fermi liquid behavior does not occur in LuV$_6$Sn$_6$. We explain the $T^{3/5}$ universal scaling behavior from the Coulomb scattering between Dirac electrons and Van Hove singularities; common features in the electronic structure of kagome materials. Finally, we show anomalous Hall-like behavior in ScV$_6$Sn$_6$ below $T_{\textrm{CDW}}$, which is absent in the Lu compound.	Comparing the transport properties of ScV$_6$Sn$_6$ and LuV$_6$Sn$_6$ is valuable to highlight the impacts of the unusual CDW in the Sc compound.\\
\end{abstract}
\begin{document}
	 \nolinenumbers
	
	\flushbottom
	\maketitle
	
	\thispagestyle{empty}

	\section*{Introduction}
	Kagome lattices comprised of appropriate constituents have the potential to host magnetism and novel topological electronic phenomena~\cite{Kag_Mag}.
	The connectivity of this lattice gives a characteristic electronic structure with Dirac points, flat bands, and van Hove singularities (VHS)~\cite{OBrien2010}. 
	Tuning the Fermi level into one of these interesting features in the band structure can induce electronic instabilities such as charge density waves (CDW), chiral spin density waves, superconductivity, and non Fermi liquid behavior~\cite{greg,Hasan,Linda2021}.
	
	The hexagonal 135 families of compounds ($A$V$_3$Sb$_5$ with $A$= K, Rb, Cs) host curious CDWs and unconventional superconductivity tied to their vanadium kagome sheets~\cite{135Cs,135K,135Rb,Ortiz2021,135_PRX,135_Nakayama2022}.
	Additionally, the normal states of the 135 families are identified as Z$_2$ topological metals with band inversion~\cite{135Cs,135_PRX}, the CDW state shows a pair density wave~\cite{135_density_wave}, nematic order~\cite{135_nematic}, and a large anomalous Hall response~\cite{135_AHE_K,135_AHE_Cs} in the absence of resolvable magnetic order~\cite{135_no_moment}. Thus there is a notable concurrence of intriguing phenomena owing to the vanadium kagome lattice.
	
	The newly discovered vanadium kagome metal, ScV$_6$Sn$_6$, also hosts a CDW below $T = $ 92 K and shares the stacked vanadium kagome sheets critical to the unique behavior in the 135 family~\cite{Sc_prl}. 
	ScV$_6$Sn$_6$ belongs to a large family of intermetallics $RT_6$$X_6$, ($R$ = rare earth; $T$ = V, Cr, Mn, Fe, Co; $X$ = Ge, Sn). Magnetism and non-trivial topological electronic properties are common features among this family ~\cite{nirmal,Madhogaria2023,GdV6Sn6_Ganesh,GdV6Sn6,TbV6Sn6_Ganesh,TbV6Sn6_ferro,Lu_Canada,GdV6Sn6_Hu}. 
	Curiously, the isoelectronic LuV$_6$Sn$_6$ and YV$_6$Sn$_6$ do not develop the CDW observed in ScV$_6$Sn$_6$~\cite{Lu_Canada,Meier2023}.
	
	The CDW in ScV$_6$Sn$_6$ has some unique characteristics. There are strong CDW fluctuations above the CDW transition with a different period than the low temperature ordered phase~\cite{ScV6Sn6_instability} demonstrating competing CDW modes~\cite{ScV6Sn6_lattice_instability}. The magnetic susceptibility~\cite{Sc_prl} and optical conductivity~\cite{ScV6Sn6_optical} reveal a drop in the number of carriers on cooling through $T_\mathrm{CDW}$. Despite this, resistivity drops through the transition due to a decreased electronic scattering by CDW fluctuations~\cite{ScV6Sn6_optical}. Although the Fermi level is dominated by $d$-orbitals from the vanadium kagome nets, the charge density wave has little impact on V positions \cite{Sc_prl} or their bands~\cite{CDW_dynamics,ScV6Sn6_spectral,ScV6Sn6_ARPES,ScV6Sn6_phonon}. In fact, the most prominent CDW gap appears in a band associated with Sn orbitals~\cite{ScV6Sn6_Sn}, highlighting the involvement of Sn in the CDW \cite{Meier2023}. Finally, Ref.~\cite{Guguchia2023} argues that the CDW in ScV$_6$Sn$_6$ breaks time-reversal symmetry based on muon spin rotation measurements. Most of the the enticing phenomena observed in the kagome compounds are linked to the proximity of their VHSs to the Fermi level; common features in the electronic structure which drive correlations. Another hallmark of kagome metals is a universal long-range Coulomb behavior\cite{DiSante2023}. Our observation of non-Fermi liquid behavior in the 135 and 166 kagome metals, which is tied to the Coulomb interactions adds another intriguing phenomena to the list of peculiar properties in these family of compounds.  
	
	We undertake a detailed study on electrical transport and magnetic properties of ScV$_6$Sn$_6$ and its sister compound LuV$_6$Sn$_6$. Although their electronic structures are nearly identical, our results suggest the Lu compound does not host a CDW.
	We compare these isostructural compounds to explore how unusual CDW order and fluctuation impact the transport properties ScV$_6$Sn$_6$ uncovering three key observations. First, both above and below the transition temperature ScV$_6$Sn$_6$ have a remarkably large magnetoresistance for current along the $c$ axis.  Second, we observe a $T^{3/5}$ non-Fermi liquid temperature dependence of the resistivity in ScV$_6$Sn$_6$ above $T_\mathrm{CDW}$, which is absent in CDW-free LuV$_6$Sn$_6$. The scaling of resistivity by $T^{3/5}$ is also observed in KV$_3$Sb$_5$, RbV$_3$Sb$_5$, and CsV$_3$Sb$_5$ suggesting this sublinear resistivity is a universal phenomenon in the vanadium CDW kagome metals above $T_\mathrm{CDW}$. We attribute the $\frac{3}{5}$ non-trivial exponent to the Coulomb scattering between fast Dirac pocket and flat Van Hove singularity pocket. Third, we note that ScV$_6$Sn$_6$ exhibits anomalous Hall effect-like behavior below $T_\mathrm{CDW}$ that is absent in the Lu compound. Our study establishes comparisons of ScV$_6$Sn$_6$ and LuV$_6$Sn$_6$ as a critical tool in future understandings of the nature of the CDW in ScV$_6$Sn$_6$.
	
	
	\section{Results and Discussion}
	\subsection{Band structures}
	\begin{figure}[!hbt]
		\centering
		\includegraphics[width=1.00\linewidth]{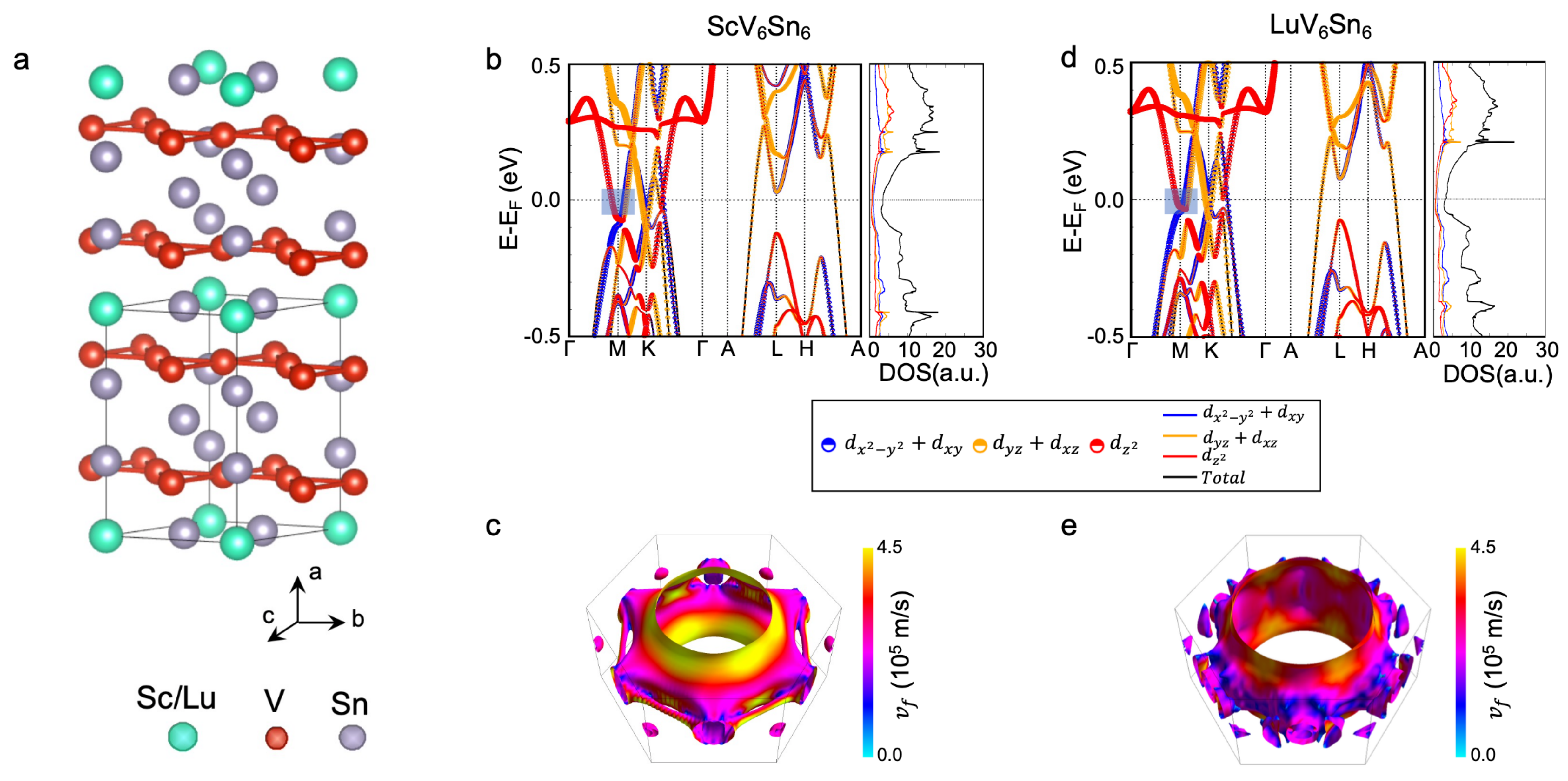}
		\centering
		\caption{Crystal and electronic structures of kagome metals ScV$_6$Sn$_6$ and LuV$_6$Sn$_6$. (a)  Crystal structure of $R$V$_6$Sn$_6$, $R$ = Sc and Lu. 
			(b) Orbital-projected band structure and density of states with the V-\textit{d} orbitals for ScV$_6$Sn$_6$ and (d) for LuV$_6$Sn$_6$. (c,e) the Fermi surface of ScV$_6$Sn$_6$ and LuV$_6$Sn$_6$, respectively, colored according to the Fermi velocity of the bands.} \label{Bands}
	\end{figure}
	The compounds ScV$_6$Sn$_6$ and LuV$_6$Sn$_6$ share numerous features, as demonstrated by their extremely similar orbital-projected electronic band structures. Specifically, both compounds exhibit a kagome lattice with a $d_{z^2}$ orbital configuration that displays a Dirac point at the K point situated below the Fermi level ($E_F$), a saddle point near $E_F$ and a flat band positioned above $E_F$ level by 0.3 eV throughout the entire Brillouin zone in the $k_z = 0$ plane. The Dirac point at the K point at -0.4 eV has $d_{x^2-y^2}+d_{xy}$ orbital character. All of these features are depicted in Fig.~\ref{Bands}(b,d). 
	
\begin{figure}[!hbt]
		\centering
		\includegraphics[width=0.55\linewidth]{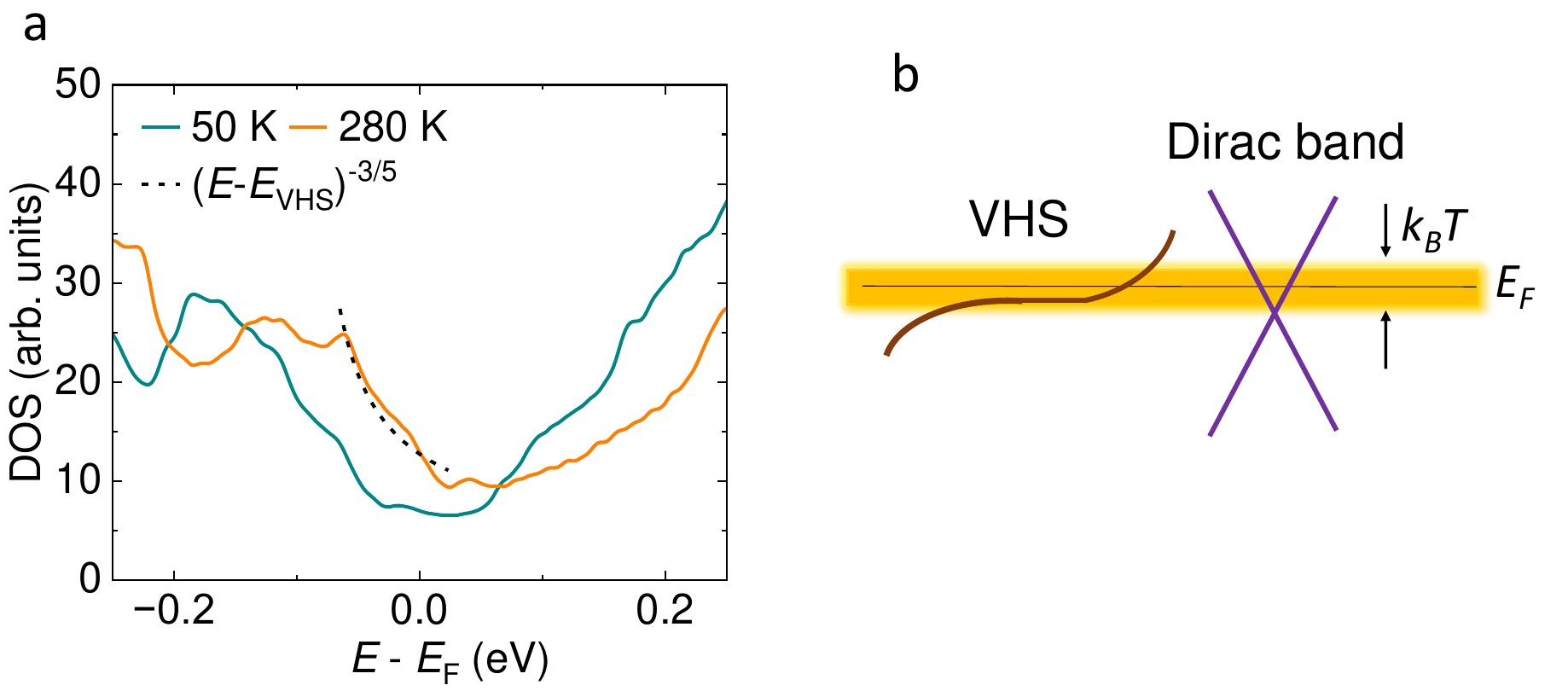}
		\centering
		\caption{Position of the saddle point below and above the CDW phase for ScV$_6$Sn$_6$. 
			(a) Density of states (DOS) plot for ScV$_6$Sn$_6$ at $T=50$ K and $T=280$ K,  with the same supercell volume given by the CDW structure. For the high-temperature structure, we fit the density of states near $E_F$ in the power law function as $(E-E_{VHS})^{-\frac{3}{5}}$. (b) The schematic diagram for the Dirac bands and the VHS near $E_F$. The yellow-shaded region represents the thermally active region in momentum space for the fast Dirac electrons and slow electrons of the VHS. 
		} \label{DOS_VHS}
	\end{figure}

	Notably, the band characterized by the $d_{z^2}$ orbital component is non-continuous along the M--K line due to interactions with non-orthogonal $d_{xz}$ and $d_{yz}$ orbital bands. Upon closer inspection, however, ScV$_6$Sn$_6$ exhibits smaller-sized electron pockets near the M-points highlighted in Fig.~\ref{Bands}(b,d) relative to LuV$_6$Sn$_6$. These diminutive pockets may facilitate superior carrier mobility, which enhances electrical conductivity. 
	The Fermi surface of ScV$_6$Sn$_6$ displays a higher Fermi velocity than LuV$_6$Sn$_6$ nearly everywhere throughout the Brillouin zone, especially near $k_{z}$=0.5 by a factor of 1.5 [Fig.~\ref{Bands}(c,e)], which enhances the carrier mobility. 
	Besides the nearly flat band around 300 meV above $E_F$, there also exists significant van Hove singularities at the hole side in ScV$_6$Sn$_6$. The band structure of ScV$_6$Sn$_6$ below the CDW phase is presented in Ref.~\cite{ScV6Sn6_ARPES}. We present the calculated DOS in Fig.~\ref{DOS_VHS}(a). We found that in ScV$_6$Sn$_6$ the VHS peak is very close to the Fermi level, $E_{\textrm{VHS}}=-80$ meV for $T>T_{\textrm{CDW}}$. In the CDW state, the VHS moves away to $E_{\textrm{VHS}}=-165$ meV. 
	This departure of the VHS away from $E_F$ leads to modification of the magnetic susceptibility and resistivity, which we explain in Sec.~\ref{chi} and Sec.~\ref{NFL}. 
	
	\subsection{Magnetic susceptibility and electrical resistivity\label{chi}}
	Sc and Lu are both nonmagnetic trivalent rare-earth elements, so we expect strong chemical and physical similarities. Although the ionic radius of Lu$^{+3}$ is about 12.3\% larger than Sc$^{+3}$~\cite{Shannon}, this only produces a subtle 1\% increase in unit cell volume. The refined structural parameters from our XRD measurements which are listed in Table~\ref{table_size} are consistent with our DFT result and those in previous works~\cite{Lu_ICSD}. The similar chemistry of Sc and Lu is also revealed by their nearly identical electronic structures (Fig.~\ref{Bands}). 
	\begin{table}[hbt]
		\centering
		\caption{\label{table_size}
			Lattice parameters and $R^{3+}$ ionic radius~\cite{Shannon} for $R$V$_6$Sn$_6$, $R$ = \{Sc,Lu\}, and charge ordering temperature. The experimental lattice parameters are obtained from the Rietveld refinement and DFT in this study.}
		\begin{tabular}{ccccc}
			& Ionic radius (\AA) & $a$ (\AA) & $c$ (\AA) & $T_{\textrm{CDW}} (K)$\\
			\hline
			ScV$_6$Sn$_6$ & 0.870  & 5.47488(6) & 9.1764(1) & 91 \\
			DFT & --  & 5.45304 & 9.23111 & -- \\
			LuV$_6$Sn$_6$ & 0.977 & 5.50286(5) & 9.1738(1) & no CDW  \\
			DFT & --  & 5.48307 & 9.22634 & -- \\ 
		\end{tabular}
	\end{table}
	
\begin{figure}[hbt]
		\centering
		\includegraphics[width=\linewidth]{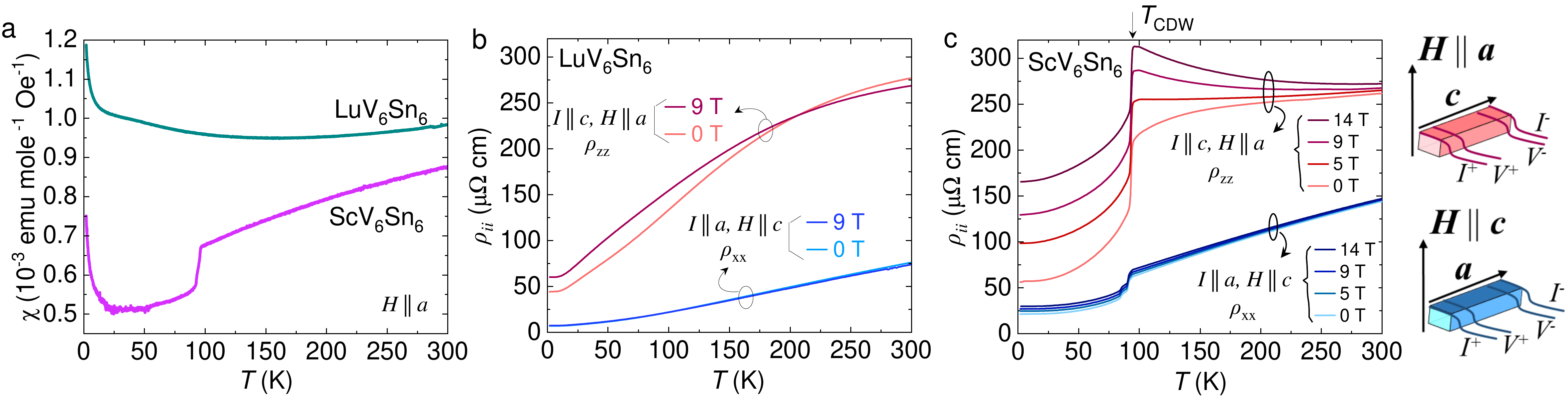}
		\centering
		\caption{Magneto-transport properties of $R$V$_6$Sn$_6$. (a) Temperature dependence of magnetic susceptibility $\chi$ for ScV$_6$Sn$_6$ and LuV$_6$Sn$_6$ with the magnetic field pointing in the plane. Electrical resistivity ($\rho$) of (b) ScV$_6$Sn$_6$ and (c) LuV$_6$Sn$_6$ as a function of temperature $T$ under the application of magnetic field $\mu_0H$. The resistivity was measured along the \textit{a} axis ($\rho_{xx}$) and \textit{c} axis ($\rho_{zz}$) of the crystals. Schematics adjacent to panel (b) show the geometry of the measurements. $T_{\textrm{CDW}}$ shows the temperature of the onset of charge ordering. Metal-like to insulator-like behavior through the application of high field is seen above the CDW state in the Sc compound.} \label{Lu_Sc_R-T_H}
	\end{figure}

	Figure~\ref{Lu_Sc_R-T_H}(a) presents the magnetic susceptibility $\chi$ as a function of temperature $T$ for the Sc and Lu compounds. Note that the magnitude has similar small values [about $1\times 10^{-3}$ emu (mol $R$V$_6$Sn$_6$)$^{-1}$ Oe$^{-1}$ or $7.7\times 10^{-5}$ cm$^3$ (mol atoms)$^{-1}$] consistent with weak Pauli paramagnetism from metals without strongly magnetic atoms~\cite{Kittel}. 
	Importantly, $\chi(T)$ of ScV$_6$Sn$_6$ drops at $T_{\textrm{CDW}} = 91$~K consistent with the CDW observed previously~\cite{Sc_prl}. 
	This is absent in the Lu material, suggesting it does not develop a CDW. Finally, note that the Lu material has a near temperature independent $\chi(T)$, in agreement with Ref.~\cite{Lu_Canada}, while ScV$_6$Sn$_6$ has a decreasing $\chi(T)$ on cooling from room temperature to $T$ = $T_{\textrm{CDW}}$. 
	The electronic contribution to the magnetic susceptibility reflects the DOS near the Fermi level with thermal broadening~\cite{Ashcroft}: 
	\begin{equation}
		\chi(T) = \mu_B^2 \mu_m \int dE D(E) \left( -\frac{df_0(E)}{dE} \right),
	\end{equation}
	where $\mu_B$ is the Bohr magneton, $\mu_m$ is the permeability, $D(E)$ is the DOS, $f_0(E) = [e^{E/(k_BT)}+1]^{-1}$ is the Fermi-Dirac distribution function, and $k_B$ is the Boltzmann constant. The $\frac{df_0(E)}{dE}$ term produces susceptibility contributions from DOS features near the Fermi level at elevated temperatures~\cite{CoSn_Meier}.
	Therefore, we attribute the decrease in $\chi(T)$ for ScV$_6$Sn$_6$ during cooling for temperatures above $T_{\textrm{CDW}}$ to the loss of thermally excited carriers from the van Hove feature just below $E_\mathrm{F}$.
	Figures~\ref{Lu_Sc_R-T_H}(b) and (c) show the in-plane longitudinal resistivity $\rho_{xx}$ and out-of-plane longitudinal resistivity $\rho_{zz}$ under the application of magnetic field for ScV$_6$Sn$_6$ and LuV$_6$Sn$_6$ samples. 
	In all the measurements, the magnetic field $\mu_0H$ was applied perpendicular to the current as depicted by the schematic in Fig.~\ref{Lu_Sc_R-T_H}(c). 
	Both samples exhibit strong resistive anisotropy expected for layered structures with $\rho_{zz}~>~\rho_{xx}$.
	This is also consistent with the Fermi velocities in-plane and out-of-plane (Fig.~\ref{Bands}). 
	In Fig.~\ref{Lu_Sc_R-T_H}(c), the CDW transition in ScV$_6$Sn$_6$ is clearly evident as a sharp drop in the resistance on cooling for both current directions ($\rho_{xx}$ drops 48.5\% and $\rho_{zz}$ drops 43.3\%). Note that the CDW transition in $\rho_{xx}$ shows a double transition (seen by others~\cite{ScV6Sn6_optical,ScV6Sn6_phonon}). The onset of the first transition, which happens at $T\sim$~94~K, coincides with the single CDW transition in $\rho_{zz}$. The midpoint of the first transition happens at $T_{\mathrm{CDW}}$ = 91~K. The onset of the second step in $\rho_{xx}$ happens at $T\sim$~89~K. We currently do not have a clear explanation for the origin of the second transition, but it might signal a lock-in transition from an incommensurate to a commensurate CDW~\cite{CDW_com_incom,CDW_theory,CDW_mohamad}. 
	There is evidence that a phonon mode with a different period is active above $T_{\mathrm{CDW}}$~\cite{ScV6Sn6_Bernevig, ScV6Sn6_instability}. 
	The dramatic drops in resistance are absent in LuV$_6$Sn$_6$ again, suggesting it does not develop charge order. Our resistivity measurement for the Lu compound is consistent with that reported in Ref.~\cite{Lu_Canada}.
	We now turn to the evolution of the resistivity under differing magnetic field strengths in these two compounds. The CDW transition temperature in ScV$_6$Sn$_6$ is insensitive to magnetic fields up to 14 T. In both compounds, the applied magnetic field has a stronger effect on $\rho_{zz}$ than $\rho_{xx}$, but the behavior in ScV$_6$Sn$_6$ is remarkable. Note that $\rho_{zz}$ rises dramatically with an increasing field below 200 K. The magnitude of this increase is nearly temperature-independent below $T_\textrm{CDW}$ and decreases slowly as the sample is heated above the transition. Curiously, above $T_\textrm{CDW}$ the slope of $\rho_{zz}(T)$ changes from positive (metal-like) to negative (insulator-like) as the field is increased through 5~T. This temperature-dependent enhancement of $\rho_{zz}$ above $T_{\mathrm{CDW}}$ may originate from strong CDW fluctuations that are recently reported in Ref.~\cite{ScV6Sn6_instability}. This dramatic magnetoresistance behavior is clearly absent in LuV$_6$Sn$_6$, revealing the anisotropic impact of CDW order and fluctuations on transport in ScV$_6$Sn$_6$. 
	
	\subsection{Non-Fermi liquid to Fermi liquid transition\label{NFL}}
\begin{figure}[!hbt]
		\centering
		\includegraphics[width=0.45\linewidth]{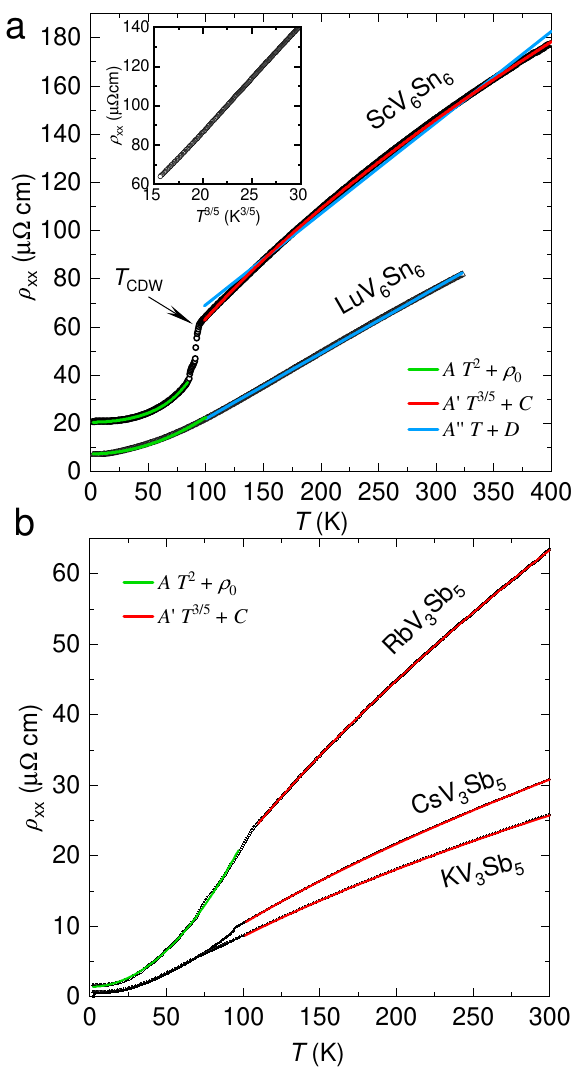}
		\centering
		\caption{Resistivity data analysis. Symbols show electrical resistivity of ScV$_6$Sn$_6$  and LuV$_6$Sn$_6$ in (a) and $A$V$_3$Sb$_5$ ($A$~=~K, Rb, Cs) in (b) as a function of temperature, with the electric current applied along the \textit{a} axis of the crystal. Red and green solid lines are fit using equations $\rho= A^\prime T^{3/5}+C$ and $\rho=A T^{2}+\rho_0$, respectively. The blue line is a linear fit to the data. $A$, $A^\prime$, $A^{\prime\prime}$, $C$, $D$, and $\rho_0$ are constant. Inset in (a) shows the resistivity above $T_\textrm{CDW}$ for ScV$_6$Sn$_6$ plotted as the $\rho~(T^{3/5})$ dependence, indicating a $T^{3/5}$ behavior at high temperatures. Data for the $A$V$_3$Sb$_5$ materials was digitized from Ref.~\cite{Rb_2nd_ref} and the authors of Refs.~\cite{135K,135Cs}.} \label{Sc_135_NF}
	\end{figure}

	Now we return to the electrical transport properties, with emphasis on $\rho_{xx}$. Figure~\ref{Sc_135_NF} shows the temperature dependence of the longitudinal resistivity $\rho_{xx}$ for the current applied along the \textit{a} axis of the crystal. Above the charge ordering temperature, $\rho_{xx}$ varies sublinearly with temperature. This sublinear behavior is a common feature in most of the kagome metallic compounds~\cite{135K,135Rb,135Cs,CoSn_Meier,LaRu3Si2_exp}.
	
	The resistivity versus temperature curves for common metallic materials generally has positive concavity. 
	Temperature dependence of a metallic resistance at zero fields can be modeled by a combination of the power law temperature dependencies $\sum_{i} \alpha_i T^{\gamma_i}$, where different terms describe different $T$-dependence scattering phenomena~\cite{sublinear}. 
	At high temperatures, the scattering is dominated by phonons, and resistivity grows linearly until the temperature approaches the Debye temperature $T_D$~\cite{Das_Sarma_T}. Our estimate for the Debye temperature from the heat capacity result reported for ScV$_6$Sn$_6$~\cite{Sc_prl} is $T_D = 316$ K. Therefore, one would expect a linear trend for $\rho_{xx} (T)$ at an intermediate and high temperature within the context of the present experiment. This is indeed the case for LuV$_6$Sn$_6$ in which the CDW transition is absent. However, the $\rho_{xx}~(T)$ curve for ScV$_6$Sn$_6$ deviates from this linear behavior. 
	We modeled the resistivity data to $\rho\propto T^n$ and found that the power-law formula with $n=0.62 \approx 3/5$ fits the experimental data extremely well for the temperature range of 100 -- 400 K, as demonstrated by the red line in Fig.~\ref{Sc_135_NF}(a). For the non-CDW compound, LuV$_6$Sn$_6$, $n=0.95\approx1$ models the resistivity data very well for the measured temperature range of 100-300 K. 
		
	We performed a similar analysis ($\rho\propto T^n$) on the resistivity of the 135 vanadium kagome metals and found that $n = 3/5$ also fits the resistivity behavior in this vanadium kagome family. The obtained fitting results are $n=0.61$ for RbV$_3$Sb$_5$, $n=0.60$ for CsV$_3$Sb$_5$, and $n=0.67$ for KV$_3$Sb$_5$. We noticed some variations between samples that are measured in different groups, and so within the uncertainty of the experimental fit, we take $n \approx 3/5$. 
	This shows that resistivity varies by $T^{\frac{3}{5}}$ for vanadium kagome compounds exhibiting CDW. In Fig.~\ref{Sc_135_NF}(b) we show that the exponent $n = 3/5$ describes the $\rho (T)$ behavior of KV$_3$Sb$_5$~\cite{135K}, RbV$_3$Sb$_5$~\cite{135Rb}, and CsV$_3$Sb$_5$~\cite{135Cs} samples very well above the charge ordering temperature. 
	
	Analysis of $\rho_{xx}$ displays that below the CDW transition, resistivity follows the $T^{2}$ behavior, as shown by the green curves in Figs.~\ref{Sc_135_NF}(a,b). Similar to ScV$_6$Sn$_6$, the $T^{2}$ scaling holds below the CDW transition for the 135 families as well. 
	
	A $T^2$ scaling of the resistivity is a signature of the Fermi liquid (FL) behavior in metals. Generally, any deviation from this scaling, such as the linear in $T$ resistivity of high-$T_c$ cuprates~\cite{NFL_SC_Taillefer} and heavy-fermion compounds~\cite{HF_CeCoIn5}
	, is loosely defined as non-Fermi liquid (non-FL) behavior~\cite{Das_Sarma_T}. We adopt the same terminology and identify the sublinear scaling of $\rho_{xx}(T)$ as non-FL behavior.  
	
	As shown in Fig.~\ref{DOS_VHS}(a), above the CDW state, the VHS is close to the Fermi level, and it moves away from $E_F$ at $T < T_{\textrm{CDW}}$. The close proximity of the VHS to $E_F$ is likely to act as the origin of the sublinear resistivity at high temperatures. Here we propose a simple theory that considers the scattering between fast Dirac and slow electrons from the flat VHS to explain this sublinear temperature dependence of $\rho_{xx}(T)$. 
	
	As indicated by the dashed curve in Fig.~\ref{DOS_VHS}(a), we fit the chemical potential dependent density of states of the VHS from the right to $(E-E_{\textrm{VHS}})^{-\frac{3}{5}}$, which grows faster than the ordinary VHS with logarithmic divergence. The electron interaction mediated by this high-order VHS leads to a non-FL behavior~\cite{supermetal}. Here we briefly show that the VHS is crucial for the transport properties, as the electron's group velocity rapidly decreases close to the singular point. Charge carriers are dominated by the fast electrons supplied by other Fermi surfaces, which are Dirac bands in ScV$_6$Sn$_6$. The transport lifetime is given by $\frac{1}{\tau}=\int \frac{d^2 \boldsymbol{q}}{(2 \pi)^2} w_{\boldsymbol{k}+\boldsymbol{q}, \boldsymbol{k}}\left(1-\cos \theta_{\boldsymbol{k}+\boldsymbol{q}, \boldsymbol{k}}\right)$. Here, $w_{\boldsymbol{k}+\boldsymbol{q}, \boldsymbol{k}}$ is the scattering rate and $\theta_{\boldsymbol{k}+\boldsymbol{q}, \boldsymbol{k}}$ is the angle between states at $\boldsymbol{k}$ and $\boldsymbol{k}+\boldsymbol{q}$~\cite{Mahan}. We consider ScV$_6$Sn$_6$ to be a quasi-2D metal with large anisotropy in band dispersion and resistivity as shown in Fig. \ref{Lu_Sc_R-T_H}(b).

	In this context, the main sources of charge carriers scattering are impurity and electron-electron scattering. Electron-electron scattering of fast Dirac electrons could happen either at other electrons from the Dirac cone (`intranode' processes) or at slow particles from the vicinity of VHS (`internode' processes). Simplest estimate for intranode scattering time $\tau_\mathrm{e-e}$ in systems with linear dispersion \cite{nonfermi1,nonfermi2} is  $\tau_\mathrm{e-e}\sim\frac{\hbar}{(e^2/\hbar v_F)^2 k_BT}\sim 10^{-16}\, s$, much smaller than typical $\tau_\mathrm{imp}$. It is important to note that internode scattering processes should have an even higher relaxation rate due to the divergent density of states near VHS. Therefore, one could see that although electron-electron scattering is momentum conserving, it should give the leading $T$-dependent contribution to resistivity $\rho(T)$. The leading contribution to the electron-electron scattering rate is provided by the scattering of fast Dirac electrons at heavy VHS states.
	
	Since slow electrons from near the high-order VHS have low energies, they can be viewed as a momentum reservoir that relaxes the momentum (and not energy) of fast electrons. The energy dispersion of slow electrons is approximately described as $Ak^5$ near the $M$ momentum. Due to its approximate flatness, 
	the size of the momentum reservoir depends on temperature as shown in Fig.~\ref{DOS_VHS}(b). Therefore, the transferred momentum is bounded by $q < q_T\equiv\left(k_B T / A\right)^{1/5}$, with $k_B T$ as the energy range of the thermally activated states. It could be shown from experimental results that the transferred momentum $q_T\sim 10^6\, \mathrm{cm}^{-1}$ is always smaller than the initial fast electron momentum $k\sim 10^7\, \mathrm{cm}^{-1}$. Thus, in this setting, fast electrons would be scattered at small angles $\theta\sim q/k\ll 1$, which yields the following estimate for momentum integral of $\frac{1}{\tau}$:
	\begin{equation}
		\frac{1}{\tau}\sim\left(\frac{k_B T}{A}\right)^{\frac{1}{5}}\cdot\left(\frac{k_B T}{A}\right)^{\frac{2}{5}}\sim \left(\frac{k_B T}{A}\right)^{\frac{3}{5}}
		\label{scatrate}
	\end{equation}
	The first multiplier in \eqref{scatrate} comes from integration over scattered electron phase volume, the second one stands for angular-dependent part $1-\cos \theta_{\boldsymbol{k}+\boldsymbol{q}, \boldsymbol{k}}\sim \theta_{\boldsymbol{k}+\boldsymbol{q}, \boldsymbol{k}}^2$. All in all, the temperature exponent of resistivity is given by 
	\begin{equation}
		\rho(T) \sim \frac{1}{e^2 D_d v_F^2 \tau} \sim \frac{1}{\tau} \sim \left(\frac{k_B T}{A}\right)^{\frac{3}{5}}
	\end{equation}
	where nontrivial exponent of $3/5 < 1$, is directly influenced by the high-order VHS
	, assuming a constant density of fast Dirac electrons $D_d$.
	Experimentally, this picture is consistent with the temperature-dependent electron mobility and almost constant electron density shown in Fig.~S3 in the supplementary information (SI).
	
	\subsection{Hall effect and Magnetoresistivity}
	We have measured the in-plane magnetoresistance MR$_{xx}$ and Hall resistivity $\rho_{yx}$, and their out-of-plane counterparts, MR$_{zz}$, and $\rho_{yz}$, for ScV$_6$Sn$_6$ and LuV$_6$Sn$_6$. Magnetoresistance is defined as $\mathrm{MR_{ii}=[\rho_{ii}\left(\mu_0\textit{H}\right)-\rho_{ii}\left( 0\right) ]/\rho_{ii}(0)\times100}\%$. 
	In the Hall effect measurements, $\rho_{ji}$, $i$ indicates the direction in which the electric current was applied, and $j$ indicates the direction of the measured transverse voltage. 
	
\begin{figure}[hbt]
		\includegraphics [width=1\linewidth] {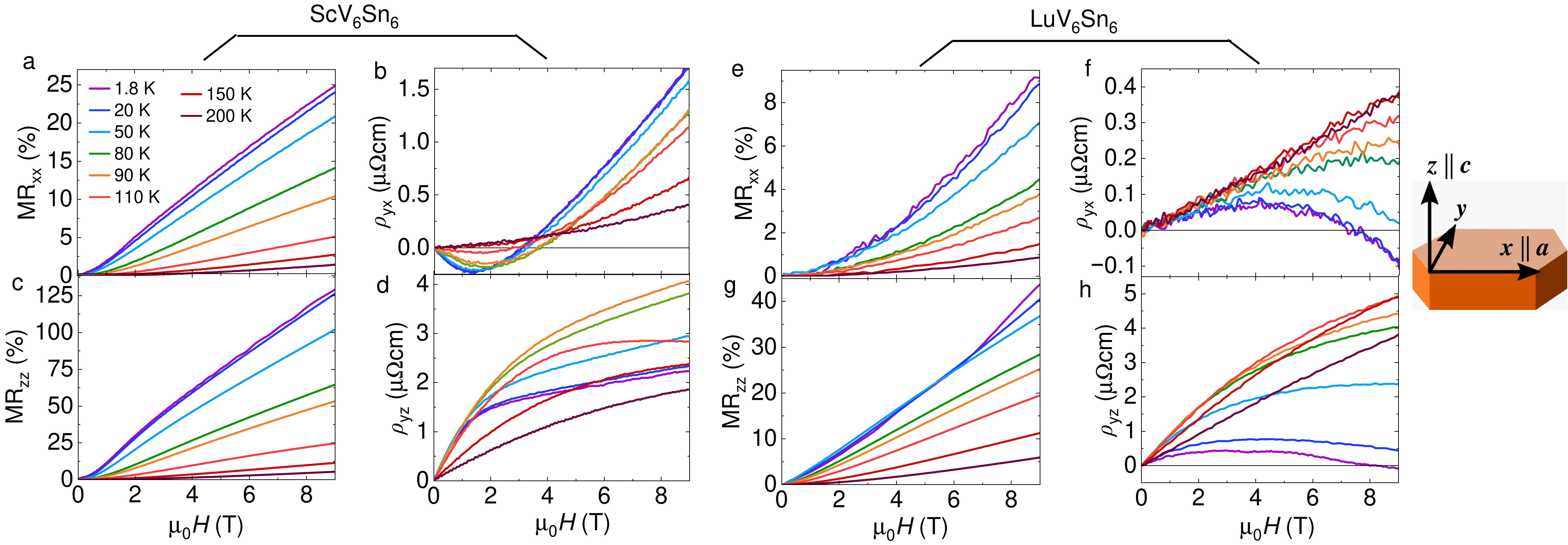}
		\caption{Comparison of magnetoresistivity (MR$_{ii}$) and Hall resistivity $\rho_{ji}$ in ScV$_6$Sn$_6$ and LuV$_6$Sn$_6$.
			(a) MR$_{xx}$ and (b) $\rho_{yx}$ as a function of magnetic field $\mu_0H$ at various temperatures when $\mu_0H~||~c$ for ScV$_6$Sn$_6$. (c) Magnetoresistance MR$_{zz}$ and (d) Hall resistivity $\rho_{yz}$ as a function of magnetic field $\mu_0H$ at various temperatures when $\mu_0H~||~a$ for ScV$_6$Sn$_6$. (e) Magnetoresistance MR$_{xx}$ and (f) Hall resistivity $\rho_{yx}$ as a function of magnetic field $\mu_0H$ at various temperatures when $\mu_0H~||~c$ for LuV$_6$Sn$_6$. (g) Magnetoresistance MR$_{zz}$ and (h) Hall resistivity $\rho_{yz}$ as a function of magnetic field $\mu_0H$ at various temperatures when $\mu_0H~||~a$ for LuV$_6$Sn$_6$. The cartoon on the right side, shows the coordinate $x,y$, and $z$ relative to the crystalographic direction.}  \label{MR_Hall}
	\end{figure}

	Fig.~\ref{MR_Hall} presents the magnetoresistance and Hall effect for ScV$_6$Sn$_6$ and LuV$_6$Sn$_6$. As pointed out above with Fig.~\ref{Lu_Sc_R-T_H}(b) and (c), magnetoresistance is largest for currents applied along the $c$ axis in both compounds. The MR$_{zz}$ of ScV$_6$Sn$_6$ is especially notable, achieving 125\% at 9 T and 1.8 K, Fig.~\ref{MR_Hall}(c). This reinforces the strong influence of the CDW order on the electronic transport and therefore transport properties of the Sc compound.

	The general trend of the Hall resistance is nonlinear in both samples, particularly below 150~K. This is consistent with the multi-band nature of these systems (Fig.~\ref{Bands}). The dominant Hall responses of $\rho_{yx}$ and $\rho_{yz}$ in LuV$_6$Sn$_6$ are qualitatively similar. However, the Hall responses in ScV$_6$Sn$_6$ are markedly different among the two measured geometries.
	In-plane Hall resistance $\rho_{yx}$ in ScV$_6$Sn$_6$ [Fig.~\ref{MR_Hall}(b)] at $T \le T_{\textrm{CDW}}$ shows a negative slope below 1.7 T and a positive slope above that. LuV$_6$Sn$_6$ also shows this nonlinear behavior but with the opposite signs. The Hall signal is observed to change sign at higher values of the field. Similar nonlinear behavior has been seen in other compounds of the 166 families, such as YV$_6$Sn$_6$ and GdV$_6$Sn$_6$~\cite{GdV6Sn6, GdV6Sn6_Ganesh}. The complex field and temperature evolution of the Hall resistivity suggests the presence of multiple carriers. One may hypothesize that the carrier with a higher (lower) concentration has a lower (higher) mobility and dominates the transport at lower (higher) fields. We fitted the in-plane magnetoresistance (MR$_{xx}$) and Hall resistivity ($\rho_{yx}$) to the simple two-band model~\cite{Mozaffari2020,2-Band} through Eq.~S1.
	The results of simultaneous fittings to Eq.~S1, charge carrier densities and mobilities as a function of temperature, are shown in Fig.~S3 in SI. For ScV$_6$Sn$_6$, we obtained electrons that are one order of magnitude less numerous than holes but have almost 3 times larger mobilities. The dominant type of carrier in LuV$_6$Sn$_6$ has an opposite sign and is more numerous compared to ScV$_6$Sn$_6$. The smaller mobility values for the Lu compound compared to the Sc compound agrees with the DFT calculation results for the Fermi velocities, shown projected on the Fermi surface [Fig.~\ref{Bands}(c,e)].

	The temperature dependence of carriers in ScV$_6$Sn$_6$ [Fig.~S3] shows a clear signature of the CDW transition. There is a drop in hole-like carriers on cooling through $T_\mathrm{CDW}$ consistent with the drop in magnetic susceptibility and opening of a CDW gap. The increase in carrier mobility during the formation of the CDW results in the observed drop in resistivity despite the less numerous carriers.

	\subsubsection{Anomalous Hall-like behavior}
	In the 135 materials, the anomalous Hall effect (AHE) has drawn significant attention. A saturating anomalous Hall-like signal is observed in the $ab$-plane, sparking speculation that the CDW breaks time-reversal symmetry~\cite{135_AHE_Cs,135_AHE_K}. Other indications of time-reversal symmetry breaking are the observation of an increase in muon spin relaxation below $T_\textrm{CDW}$~\cite{muon_1,muon_2_Nature} and magneto-optical Kerr effect~\cite{Kerr_1}. In the case of the latter experiment, there remains debate as an alternative experiment found no discernible Kerr effect~\cite{Kapitulnik}. Since long-range magnetic order is absent in the 135 compounds, time-reversal symmetry breaking is explained in terms of loop-current order in the CDW state~\cite{Kerr_1,greg}. 
	Do the vanadium kagome metals ScV$_6$Sn$_6$ or LuV$_6$Sn$_6$ share this AHE behavior?
	
\begin{figure}[!hbt]
		\centering
		\includegraphics[width=0.55\linewidth]{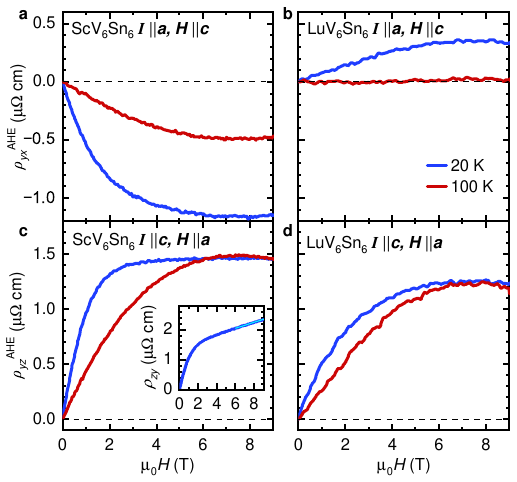}
		\centering
		\caption{
			Extracted $\rho_{yz}^{AHE}$ taken by subtracting the local linear ordinary Hall background between 6 and 9~T. 
			The inset in panel (c) illustrates the fitting (cyan line) done to estimate the normal Hall contribution.
		} \label{AHE}
	\end{figure}

	In Fig.~\ref{AHE}, we attempt to capture an AHE in ScV$_6$Sn$_6$ and LuV$_6$Sn$_6$. Our emphasis is on the $yz$ direction where the curves for ScV$_6$Sn$_6$ below the CDW transition show characteristics of AHE. 
	First, we estimate an ordinary contribution to the Hall signal by fitting the slope of $\rho_{ji}(\mu_0H)$ between 6 and 9\,T (inset in Fig.~\ref{AHE}). We then subtract to obtain $\rho_{ji}^{\mathrm{AHE}}$ to highlight abrupt changes in slope others have identified as an anomalous Hall contribution.
	
	The procedure yields smooth and kink-free curves for $\rho_{yx}^{\mathrm{AHE}}$ for both compounds at $T$  = 20 and 100 K. 
	We tie the nonlinear Hall behavior in both compounds to the multi-band Fermi surfaces.
	Only the $\rho_{yz}^{\mathrm{AHE}}$ from ScV$_6$Sn$_6$ curves below $T_{\mathrm{CDW}}$ has the distinct two-slope character [the blue 20\,K curve in Fig.~\ref{AHE}(c)]. This could be evidence of time reversal symmetry breaking in the CDW state by weak magnetism or loop-current order. 
	Ref.~\cite{Guguchia2023} states that there is no evidence of static magnetic order but suggests that time reversal symmetry might still be broken. Moreover, orbital current order has also been predicted to occur in ScV$_6$Sn$_6$~\cite{Francesco}.
	LuV$_6$Sn$_6$ does not have the CDW, and its $\rho_{yz}^{\mathrm{AHE}}(\mu_0H)$ does not show this two-slope behavior. The different Hall behavior of the two compounds is emphasized by plotting the Hall conductivity, $\sigma_{ij}$ (Fig.~S4 in SI).
	The AHE has also been reported in Ref.~\cite{ScV6Sn6_AHE}, but we provide a comparison to LuV$_6$Sn$_6$. This comparison suggests that the AHE-like behavior in ScV$_6$Sn$_6$ arises from the CDW order. 
	
	In summary, we have studied the anisotropic transport properties of two kagome compounds, ScV$_6$Sn$_6$ and LuV$_6$Sn$_6$.  While the compounds share similar band structures, the former exhibits a CDW phase at $T_{\textrm{CDW}}$ = 91 K, and the latter does not. By comparing the properties of these materials, we uncover three key phenomena about ScV$_6$Sn$_6$. First, ScV$_6$Sn$_6$ has strong magnetoresistance for current along the $c$ axis that is absent in the Lu compound. This behavior occurs not only below $T_\mathrm{CDW}$, but also extends to temperatures well above the transition suggesting a connection to CDW fluctuations. Second, we observe ScV$_6$Sn$_6$ has a sublinear temperature dependence of resistivity above $T_\mathrm{CDW}$, which scales by $T^{\frac{3}{5}}$. This non-Fermi liquid behavior also occurs in KV$_3$Sb$_5$, RbV$_3$Sb$_5$, and CsV$_3$Sb$_5$ suggesting this is a universal feature of the vanadium kagome CDW materials. Resistivity in the non-CDW compound LuV$_6$Sn$_6$ has a linear temperature dependence.	Finally, we identify AHE-like behavior in ScV$_6$Sn$_6$ below $T_\mathrm{CDW}$, which is absent in LuV$_6$Sn$_6$. This could be evidence of time reversal symmetry breaking or a consequence of reconstructing the intricate Fermi surface. 	Our comparison of magnetotransport in ScV$_6$Sn$_6$ and LuV$_6$Sn$_6$ reveals the unique influence of CDW order in the Sc material. These materials siblings will provide the foundation for understanding CDW order in the kagome metals.
	
	
	\section*{Methods}
	\textbf{Growth and characterization}: ScV$_6$Sn$_6$ and LuV$_6$Sn$_6$ were synthesized via the tin flux method with a starting atomic ratio of Sc/Lu:V:Sn = 1:6:60.
	Sc/Lu pieces (Alfa Aesar 99.9\%), V pieces (Alfa Aesar 99.8\%), and Sn shot (Alfa Aesar 99.9999\%) were loaded into an alumina Canfield crucible set.
	The crucible assembly was sealed within a fused silica ampule, heated to 1150 $^\circ$C over 12 h, held for 15 h, and cooled to 780 $^\circ$C over 300 h. At this temperature, the flux was separated from the crystals by inverting the tube and centrifuging. This procedure yielded bulky, hexagonal metallic crystals with typical lateral dimensions of 0.5--4 mm in size. 	
	The elemental composition and approximate stoichiometry of the structure were confirmed via scanning electron microscopy (SEM)/energy-dispersive spectroscopy (EDS). A powder X-ray diffraction (XRD) measurement was performed using a Bruker D2 Phaser equipped with a Cu \textit{K}$_\alpha$ X-ray source. Figures S1 and S2 in the SI confirm the \textit{P6/mmm} HfFe$_6$Ge$_6$-type structure for both samples. 
	Conventional magnetotransport experiments were performed in a physical property measurement system (Quantum Design-PPMS) under magnetic fields up to $\mathrm{\mu_0} H = 9$ T and temperatures as low as 1.8 K. The resistivity was measured with the standard four-probe method with the electrical current applied in the \textit{ab}-plane and out-of-plane. Magnetization measurements were performed in a commercial superconducting quantum interference device magnetometer (Quantum Design).
	We define $\mathbf{x} \parallel \mathbf{a}$, $\mathbf{y} \parallel \mathbf{c}\times\mathbf{a}$, and $\mathbf{z} \parallel \mathbf{c}$, which is depicted by the cartoon in Fig.~\ref{MR_Hall} \\
	\textbf{Simulations}: We performed ab \textit{initio} calculations based on density functional theory (DFT)~\cite{DFT_Hohenberg, DFT_Kohn} as implemented in the Vienna ab \textit{initio} simulation package (VASP)~\cite{ab_Kresse_1993}
	with projector augmented wave potentials~\cite{Bloechl_1994}
	and spin-orbit coupling. The Perdew-Burke-Ernzerhof (PBE) form~\cite{Perdew1996} was employed for the exchange-correlation functional with the generalized gradient approximation (GGA). The energy cutoff was set to 520 eV for all calculations. The Brillouin zone was sampled using a $31\times31\times31$ $\Gamma$-centered \textit{k}-grid. Atomic relaxations were done until the Helmann-Feynman force acting on every atom became smaller than 0.01 eV/\AA. Both ScV$_6$Sn$_6$ and LuV$_6$Sn$_6$ adopt the \textit{P6/mmm} space group, with the unit cell containing two kagome layers composed of vanadium atoms. These layers are enclosed by Sn and ScSn/LuSn layers along the out-of-plane direction in ScV$_6$Sn$_6$ and LuV$_6$Sn$_6$, respectively. As shown in Table~\ref{table_size}, the DFT lattice parameters of ScV$_6$Sn$_6$ are found to be $a = 5.45403$~\AA~and $c = 9.23111$~\AA, while those of LuV$_6$Sn$_6$ are $a = 5.50286$\,\AA~and~$c = 9.22634$\,\AA. These values are in agreement with experimental measurements, which show that the $a$ lattice parameter of ScV$_6$Sn$_6$ is smaller than that of LuV$_6$Sn$_6$, and the $c$ lattice parameter of the former is larger than that of the latter.
	We calculate the density of states (DOS) using a tight binding model constructed from an automatic Wannierization scheme~\cite{zhang2021different} with FPLO~\cite{koepernik1999full}. The momentum space integration mesh is $100\times100\times 100$ to account for the fine variations around the Fermi level, which can be important for comparison with the experimentally measured magnetic susceptibility and resistivity.
	

	\section*{Acknowledgements }
	S.~M., R.~P.~M., and D.~M. acknowledge the support from AFOSR MURI (Novel Light-Matter Interactions in Topologically Non-Trivial Weyl Semimetal Structures and Systems), grant\# FA9550-20-1-0322. W.~R.~M. and H.~W.~S.~A. acknowledge the support from the Gordon and Betty Moore Foundation's EPiQS Initiative, Grant GBMF9069 to D.~M. Theory work by Y.~Z. was supported from the start-up fund at the University of Tennessee. Theory work at the Oak Ridge was supported by the U.S. Department of Energy (DOE), Office of Science, National Quantum Information Science Research Centers, Quantum Science Center (S.-H.~K.), and by the U.S. Department of Energy, Office of Science, Office of Basic Energy Sciences, Materials Sciences and Engineering Division (J.~W.~V. and M.~Y.). This research used resources of the Oak Ridge Leadership Computing Facility at the Oak Ridge National Laboratory, which is supported by the Office of Science of the U.S. Department of Energy under Contract No. DE-AC05-00OR22725 and resources of the National Energy Research Scientific Computing Center, a DOE Office of Science User Facility supported by the Office of Science of the U.S. Department of Energy under Contract No. DE-AC02-05CH11231 using NERSC award BES-ERCAP0024568. L.~L. acknowledges the support from the National Science Foundation under Award No. DMR-2004288 and the Department of Energy under Award No. DE-SC0020184.\\
	We thank Brenden Ortiz and Andrea Capa Slinas for sharing the raw data for KV$_3$Sb$_5$ and CsV$_3$Sb$_5$ resistivity.

	\section*{Author contributions statement}
	S.~M. and D.~M. conceived of the study. S.~M., W.~R.~M., R.~P.~M., and H.~W.~S.~A. grew the crystals. S.~M., W.~R.~M., and R.~P.~M. performed the magneto-transport measurements and analyzed the obtained data. G.~Z., Y.~Z., K.-W~C., K.~J., D.~Z., A.~C., and L.~L. were involved in the discussion of the data. 
	S.-H.~K., J.~W.~V., and M.~Y. performed the DFT calculations. Y.~Z. and N.~P. provided the theoretical analysis of the scaling of the resistivity. S.~M. wrote the manuscript with inputs from all the authors. 
	
	\section*{Competing interests}
	All other authors declare they have no competing interests. 
	
	\section*{Corresponding authors}
	Correspondence to smozaff1@utk.edu and dmandrus@utk.edu
	
\end{document}